# On Tractable Exponential Sums

Jin-Yi Cai [*]     Xi Chen [†]     Richard Lipton [‡]     Pinyan Lu [§]

**Abstract**

We consider the problem of evaluating certain exponential sums. These sums take the form

$$\sum_{x_1, x_2, \ldots, x_n \in \mathbb{Z}_N} e^{\frac{2\pi i}{N} f(x_1, x_2, \ldots, x_n)},$$

where each $x_i$ is summed over a ring $\mathbb{Z}_N$, and $f(x_1, x_2, \ldots, x_n)$ is a multivariate polynomial with integer coefficients. We show that the sum can be evaluated in polynomial time in $n$ and $\log N$ when $f$ is a quadratic polynomial. This is true even when the factorization of $N$ is unknown. Previously, this was known for a prime modulus $N$. On the other hand, for very specific families of polynomials of degree $\geq 3$ we show the problem is #P-hard, even for any fixed prime or prime power modulus. This leads to a complexity dichotomy theorem — a complete classification of each problem to be either computable in polynomial time or #P-hard — for a class of exponential sums. These sums arise in the classifications of graph homomorphisms and some other counting CSP type problems, and these results lead to complexity dichotomy theorems. For the polynomial-time algorithm, Gauss sums form the basic building blocks. For the hardness results, we prove group-theoretic necessary conditions for tractability. These tests imply that the problem is #P-hard for even very restricted families of simple cubic polynomials over fixed modulus $N$.

## 1 Introduction

Exponential sums are among the most studied objects in number theory [12, 10, 11]. They have fascinating properties and innumerable applications. Recently they have also played a pivotal role in the study of the complexity of graph homomorphisms [8, 2].

The most fundamental and well-known among exponential sums are those named after Gauss. The simplest Gauss sum is as follows. Let $p$ be an odd prime, and $\omega_p = e^{2\pi i/p}$ be the $p$-th primitive root of unity. Then the Gauss sum over $\mathbb{Z}_p$ is

$$G = \sum_{t \in \mathbb{Z}_p} \left(\frac{t}{p}\right) \omega_p^t, \quad \text{where } \left(\frac{t}{p}\right) \text{ is the Legendre symbol.} \tag{1}$$

---
[*]University of Wisconsin-Madison, `jyc@cs.wisc.edu`
[†]University of Southern California, `csxichen@gmail.com`
[‡]Georgia Institute of Technology, `richard.lipton@cc.gatech.edu`
[§]Microsoft Research Asia, `pinyanl@microsoft.com`

In this paper, we will need to use a more general form of the Gauss sum which will be defined later in Section 2. Another well-known expression for $G$ in (1) is:

$$G = \sum_{x \in \mathbb{Z}_p} (\omega_p)^{x^2},$$

where the exponents are quadratic residues (mod $p$), and one can observe that $(\omega_p)^{x^2}$ are somewhat "randomly" distributed on the unit circle. Gauss knew the remarkable equality $G^2 = (-1)^{(p-1)/2}p$:

$$G = \pm\sqrt{p} \quad \text{if } p \equiv 1 \pmod{4}, \quad \text{and} \quad G = \pm i\sqrt{p} \quad \text{if } p \equiv 3 \pmod{4}. \tag{2}$$

In particular $|G| = \sqrt{p}$ which is an expression that the $p$ terms in the sum $G$ are somewhat "random" (but note that the equality is exact). However, the truly amazing fact is that, in all cases, the plus sign (+) always holds in (2). Gauss recorded this conjecture in his diary in May 1801, and on August 30, 1805, Gauss recorded that a proof of the "very elegant theorem mentioned in 1801" had finally been achieved.[1]

In this paper, we consider the complexity of evaluating exponential sums of the form

$$Z(N, f) = \sum_{x_1, \ldots, x_n \in \mathbb{Z}_N} e^{\frac{2\pi i}{N} f(x_1, \ldots, x_n)},$$

where each $x_i$ is summed over a ring $\mathbb{Z}_N$ and $f(x_1, \ldots, x_n)$ is a multivariate polynomial with integer coefficients. (We may assume that the coefficients of $f$ are from $\mathbb{Z}_N$.) The output of this computation is an algebraic number, in the cyclotomic field $\mathbb{Q}(e^{2\pi i/N})$. Any canonical representation of the output algebraic number will be acceptable [9, 7]. These sums are natural generalizations of the sums considered by Gauss, and with arbitrary polynomials $f(x_1, x_2, \ldots, x_n)$, they have also played important roles in the development of number theory.

Our main results are as follows: We show that the sum $Z(N, f)$ can be evaluated in polynomial time when $f$ is a quadratic polynomial. The computational complexity is measured in terms of $n$, $\log N$, and the number of bits needed to describe $f$. While it is known that $Z(N, f)$ can be computed efficiently when $N$ is a prime [13], our algorithm works for any composite modulus $N$, even without knowing its prime factorization. On the other hand, for very specific families of polynomials of degree $\geq 3$ we show the problem is #P-hard even for any fixed prime or prime power modulus. This leads to a complexity dichotomy theorem — a complete classification of each problem to be either computable in polynomial time or #P-hard — for a class of exponential sums.

For the polynomial-time algorithm, we employ an iterative process to eliminate one variable at a time. Gauss sums form the basic building blocks. The fact that we know the exact answer to the Gauss sum, *including the sign*, is crucial. It turns out that the situation is different for an odd or an even modulus $N$. A natural idea is to deal with each prime power in the modulus $N$ separately, and combine the answers by Chinese remaindering. It turns out that the algorithm is more difficult for a modulus which is a power of 2, than for an odd prime power. A more fundamental difficulty arises when $N$ is large and its prime factorization is unknown. We overcome this difficulty as follows: (1) Factor out all powers of 2 in $N$ and deal with it separately. (2) Operate in the remaining odd modulus *as if it were* an odd prime power; whenever this operational commingling encounters an obstacle, we manage to discover a non-trivial factorization of the modulus into relatively prime parts. In that case we recurse.

---

[1] In a letter dated September 3, 1805 Gauss wrote that ... *Seldom had a week passed for four years that he had not tried in vein to prove his conjecture of 1801*... Finally, "*Wie der Blitz einschlägt, hat sich das Räthsel gelöst* ..." ("*as lightning strikes was the puzzle solved* ...").

**Theorem 1.1.** *Let $N$ be any positive integer and $f \in \mathbb{Z}[x_1, \ldots, x_n]$ be a quadratic polynomial (every monomial has degree at most 2) in $n$ variables $x_1, \ldots, x_n$. Then the sum $Z(N, f)$ can be evaluated in polynomial time in $n$, $\log N$, and the number of bits needed to describe $f$.*

Previously it was known that for quadratic polynomials $f$ the sum can be computed in polynomial time, if $N$ is a prime [13]. An algorithm with running time $O(n^3)$ can also be found in the paper by Ehrenfeucht and Karpinski [5]. Compared to these algorithms, ours works for any $N$ even if it is given as a part of the input and its factorization is unknown. It was also suggested that there is a reduction from root counting. One can express the sum as

$$\sum_{k=0}^{N-1} \#[f = k] \cdot e^{2\pi i k / N}.$$

If $N$ is polynomially bounded and if one can compute $\#[f = k]$ for all $k$, then one can compute the sum. But this works only when $N$ is small. Our results are for general $N$ (polynomial time in the length $\log N$). In our algorithm, Gauss sums play a crucial role. Any claim to the contrary amounts to an independent proof of Gauss's sign formula (that "very elegant theorem mentioned in 1801") since it is not only a crucial building block of our algorithm, but it is also a special case of the algorithm. We also note that our treatment for the case when $N$ is a power of 2 is significantly different than previous work. No simple adaptation of ideas from Sylvester's law of inertia seems to work.

For the hardness result, we give several successively more stringent necessary conditions for a class of polynomials to be tractable. The first necessary condition involves the rank of an associated matrix, and the proof uses the widely applicable dichotomy theorem of Bulatov and Grohe [1] for counting graph homomorphisms over non-negative weighted graphs. The second necessary condition involves linear independence and orthogonality. The third and much more stringent necessary condition is group-theoretic in nature; it asserts that the set of row vectors of a certain complex matrix must form a group. In [8], Goldberg et al. had proved a similar condition for $\{-1, +1\}$ matrices, in the study of graph homomorphisms over real weighted graphs. Finally, in subsection 5.4 we give a generalized Group Condition which leads to a complexity dichotomy.

Previously, it was shown by Ehrenfeucht and Karpinski that for any fixed prime $N$, the problem of computing $Z(N, f)$ for general cubic polynomials is #P-hard [5]. However, our tests in Section 5 are more powerful. They allow us to prove the #P-hardness of $Z(N, f)$ even if $f$ belongs to some very restricted families of polynomials, since they fail one of the tests in Section 5 (see Corollary 5.2 and Corollary 5.3 for examples).

These sums arise recently in the classifications of graph homomorphisms and some other counting CSP type problems (a class of local constraint problems, including CSP as well as Holant Problems). For example, the tractability of the special case when $N = 2$ is a key component of the dichotomy theorem of Goldberg et al. [8] for graph homomorphisms over real weighted graphs. In particular, it implies that the graph homomorphism function $Z_{\mathbf{H}}(\cdot)$ (see definition in Section 5) with

$$\mathbf{H} = \begin{pmatrix} 1 & 1 \\ 1 & -1 \end{pmatrix},$$

which had been an obstacle to the dichotomy theorem of Bulatov and Grohe [1] and was left open for some time, can actually be computed in polynomial time.

## 2 Preliminaries

Let $\omega_N = e^{2\pi i/N}$ denote the $N$th primitive root of unity. Let $N = N_1 N_2$ be a non-trivial factorization namely $N_1, N_2 > 1$. Suppose $N_1$ and $N_2$ are relatively prime, $\gcd(N_1, N_2) = 1$. Then there exist two integers $a$ and $b$, such that $bN_1 + aN_2 = 1$. It follows that

$$\omega_N = e^{2\pi i a/N_1} \cdot e^{2\pi i b/N_2} = \omega_{N_1}^a \cdot \omega_{N_2}^b.$$

By Chinese remaindering, the mapping $x \mapsto (y, z)$, where $y \equiv x \pmod{N_1}$ and $z \equiv x \pmod{N_2}$, is an isomorphism from $\mathbb{Z}_N \to \mathbb{Z}_{N_1} \times \mathbb{Z}_{N_2}$. Thus, we have

**Lemma 2.1.** *If $bN_1 + aN_2 = 1$, then $Z(N, f) = Z(N_1, af) \cdot Z(N_2, bf)$.*

It follows that if we know a non-trivial factorization of $N$ into relatively prime factors $N_1$ and $N_2$, then the problem $Z(N, f)$ decomposes. In particular, we can factor $N = 2^k N'$, where $N'$ is odd. Thus we can treat the problems $Z(2^k, \cdot)$ and $Z(N', \cdot)$ separately. In Section 3, we give an algorithm for the case when $N$ is odd and in Section 4 we deal with the case when $N = 2^k$. Theorem 1.1 then follows from these algorithms and Lemma 2.1.

Our algorithm crucially relies on the fact that the following general form of Gauss sum $G(a,b)$ can be computed in polynomial time in $\log a$ and $\log b$, without knowing their prime factorizations. Let $a$ and $b$ be non-zero integers, $b > 0$, and $\gcd(a, b) = 1$. Then $G(a, b)$ denotes the following sum:

$$G(a,b) = \sum_{x \in \mathbb{Z}_b} \omega_b^{ax^2}.$$

A list of properties that we need to prove the tractability of $G(a, b)$ can be found in Appendix A.

## 3 Odd Modulus

In this section, we present a polynomial-time algorithm for the case when $N$ is odd.

**Lemma 3.1.** *Given an odd positive integer $N$ and a quadratic polynomial $f(x_1, x_2, \ldots, x_n) \in \mathbb{Z}[x_1, x_2, \ldots, x_n]$, the sum $Z(N, f)$ can be computed in polynomial time.*

*Proof.* Consider the expression

$$f(x_1, \ldots, x_n) = \sum_{i \leq j \in [n]} c_{i,j} x_i x_j + \sum_{i \in [n]} c_i x_i + c_0. \tag{3}$$

We may assume $c_0 = 0$, because it only contributes a constant factor to $Z(N, f)$. For every non-zero coefficient $c = c_{i,j}$ or $c_i$ of $f$, we first compute the greatest common divisor $g = \gcd(N, c^{\lfloor \log_2 N \rfloor})$. Note that, if $\mathrm{ord}_p N$ is the exact order of a prime $p$ in $N$, then $N \geq p^{\mathrm{ord}_p N}$ and thus $\mathrm{ord}_p N \leq \lfloor \log_2 N \rfloor$. Hence if $c$ shares any prime $p$ with $N$, but not all the prime factors of $N$, then $g$ has the factor $p^{\mathrm{ord}_p N}$ and $N = g \cdot (N/g)$ is a non-trivial factorization of $N$ into two relatively prime factors, $g$ and $N/g$. We can test for every nonzero coefficient $c = c_{i,j}$ or $c_i$ whether $N = g \cdot (N/g)$ gives us a non-trivial factorization of $N$ into two relatively prime factors.

By Lemma 2.1, if for some coefficient $c$, we did find such a factorization $N = N_1 \cdot N_2$, then the problem of computing $Z(N, \cdot)$ decomposes into two subproblems $Z(N_1, \cdot)$ and $Z(N_2, \cdot)$. There can be altogether at most a linear number $\log_2 N$ many subproblems. Therefore, a polynomial-time algorithm for each subproblem will give a polynomial-time algorithm for $Z(N, f)$.

Hence in the following we assume for each nonzero coefficient $c = c_{i,j}$ or $c_i$, either $\gcd(N, c) = 1$ or $c$ has all prime factors of $N$, and we know, by computing the gcd, which case it is for each $c$.

We consider the following four cases.

**Case 1**. There exists some diagonal coefficient $c_{i,i}$ relatively prime to $N$.

Without loss of generality, we assume $c_{1,1}$ is relatively prime to $N$.

Then $c_{1,1}$ is invertible in $\mathbb{Z}_N$. As $N$ is odd, 2 is also invertible. Denote by $c'_{1,i}$ an integer such that $c'_{1,i} \equiv (2c_{1,1})^{-1} c_{1,i} \pmod{N}$, for $2 \leq i \leq n$. We have, modulo $N$,

$$f(x_1, \ldots, x_n) = c_{1,1}\left[x_1^2 + 2x_1(c'_{1,2}x_2 + \ldots + c'_{1,n}x_n)\right] + \sum_{2 \leq i \leq j \leq n} c_{i,j} x_i x_j + \sum_{i \in [n]} c_i x_i$$
$$= c_{1,1}\left[x_1 + g(x_2, \ldots, x_n)\right]^2 + c_1\left[x_1 + g(x_2, \ldots, x_n)\right] + h(x_2, \ldots, x_n),$$

where $g(x_2, \ldots, x_n) = c'_{1,2} x_2 + \ldots + c'_{1,n} x_n$ and $h$ is some quadratic polynomial in $x_2, \ldots, x_n$.

If we substitute $y = x_1 + g(x_2, \ldots, x_n)$ for $x_1$, then for any fixed $x_2, \ldots, x_n \in \mathbb{Z}_N$, when $x_1$ takes all the values in $\mathbb{Z}_N$, $y$ also takes all the values in $\mathbb{Z}_N$. Hence, we have

$$Z(N, f) = \sum_{x_2, \ldots, x_n \in \mathbb{Z}_N} \sum_{y \in \mathbb{Z}_N} \omega_N^{c_{1,1} y^2 + c_1 y + h(x_2, \ldots, x_n)}.$$

Completing the square again, $c_{1,1} y^2 + c_1 y = c_{1,1}(y + (2c_{1,1})^{-1} c_1)^2 + c'$, where $c' = -c_1^2/(4c_{1,1}) \in \mathbb{Z}_N$,

$$Z(N, f) = \sum_{x_2, \ldots, x_n \in \mathbb{Z}_N} \sum_{z \in \mathbb{Z}_N} \omega_N^{c_{1,1} z^2 + h'(x_2, \ldots, x_n)},$$

where $h'(x_2, \ldots, x_n) = h(x_2, \ldots, x_n) + c'$ is an explicitly computed quadratic polynomial in $x_2, \ldots, x_n$. It then follows that $Z(N, f) = Z(N, h') \cdot G(c_{1,1}, N)$ where $h'$ has one fewer variable than $f$ and the Gauss sum $G(c_{1,1}, N)$ can be computed in polynomial time. This completes the proof of Case 1.

**Case 2**. No $c_{i,i}$ is relatively prime to $N$, but there exist some $1 \leq i < j \leq n$ such that $\gcd(c_{i,j}, N) = 1$.

By our earlier assumption, for every prime factor $p$ of $N$, $p$ divides every $c_{i,i}$, for $1 \leq i \leq n$.

The existence of $c_{i,j}$ for some $i, j : 1 \leq i < j \leq n$ implies that in particular $n \geq 2$. Without loss of generality, we assume $\gcd(c_{1,2}, N) = 1$. Now we perform the following substitution,

$$x_1 = y_1 + y_2, \quad \text{and} \quad x_2 = y_1 - y_2,$$

and $x_i$ are unchanged, for any $2 < i \leq n$ if $n > 2$. This transformation is a 1-1 correspondence from $\mathbb{Z}_N^n$ to itself with inverse $y_1 = (x_1 + x_2)/2$ and $y_2 = (x_1 - x_2)/2$, because 2 is invertible in $\mathbb{Z}_N$. Since the transformation is linear it does not change the degree of $f$. It is easily checked that the coefficient of $y_1^2$ in the new polynomial is $c_{1,1} + c_{2,2} + c_{1,2}$. Since $c_{1,1}$ and $c_{2,2}$ have all the prime factors of $N$, $c_{1,1} + c_{2,2} + c_{1,2}$ is relatively prime to $N$. As a result, this transformation reduces the computation of $Z(N, f)$ to Case 1.

**Case 3.** No coefficients $c_{i,j}$, where $1 \leq i \leq j \leq n$, are relatively prime to $N$. However, there exists a $c_i$ relatively prime to $N$, for some $1 \leq i \leq n$.

Without loss of generality, assume $\gcd(c_1, N) = 1$. Let $p$ be a prime divisor of $N$, then

$$p \mid c_{1,1}, \ldots, c_{1,n} \quad \text{and yet} \quad p \nmid c_1. \tag{4}$$

Let $k = \operatorname{ord}_p N$ be the exact order of $p$ in $N$. Then $k \geq 1$. Write $N = p^k N_1$, then $\gcd(p, N_1) = 1$, and for some integers $a$ and $b$ we have $bp^k + aN_1 = 1$. By Lemma 2.1, $Z(N, f) = Z(p^k, af) \cdot Z(N_1, bf)$. Note that $\gcd(a, p) = 1$. Hence the condition (4) above for the coefficients of $f$ also holds for $af$. We will show $Z(p^k, af) = 0$. For notational simplicity, we will write below $f$ for $af$.

$$Z(p^k, f) = \sum_{x_2, \ldots, x_n \in \mathbb{Z}_{p^k}} \omega_{p^k}^{\sum_{2 \leq i \leq j \leq n} c_{i,j} x_i x_j + \sum_{2 \leq i \leq n} c_i x_i} \sum_{x_1 \in \mathbb{Z}_{p^k}} \omega_{p^k}^{\sum_{1 \leq i \leq n} c_{1,i} x_1 x_i + c_1 x_1}.$$

We fix any $x_2, \ldots, x_n \in \mathbb{Z}_{p^k}$, and consider the inner sum over $x_1$.

If $k = 1$, then all terms $c_{1,i} x_1 x_i$ disappear, and because $p \nmid c_1$, the inner sum over $x_1$ is equal to 0.

Suppose $k > 1$. Then we repeat the sum for $p$ times with $x^{(j)} = x_1 + j \cdot p^{k-1}$, where $0 \leq j \leq p - 1$. By (4), we have $c_{1,i} x_1 x_i \equiv c_{1,i} x^{(j)} x_i \pmod{p^k}$ and

$$\sum_{x_1 \in \mathbb{Z}_{p^k}} \omega_{p^k}^{\sum_{1 \leq i \leq n} c_{1,i} x_1 x_i + c_1 x_1} = \frac{1}{p} \sum_{x_1 \in \mathbb{Z}_{p^k}} \omega_{p^k}^{\sum_{1 \leq i \leq n} c_{1,i} x_1 x_i + c_1 x_1} \left( \sum_{j=0}^{p-1} \omega_p^{jc_1} \right).$$

By $p \nmid c_1$, the geometric sum $\sum_{j=0}^{p-1} \omega_p^{jc_1} = 0$. This finishes Case 3.

**Case 4.** No coefficients $c_{i,j}$ and $c_\ell$, where $1 \leq i \leq j \leq n$ and $1 \leq \ell \leq n$, are relatively prime to $N$.

By our earlier assumption, this means that every prime factor of $N$ divides every coefficient $c_{i,j}$ and $c_\ell$. Then we can find the joint gcd $d$ of $N$ with all these coefficients, which must at least contain every prime factor of $N$, and divide out $d$ in the exponent. By $\omega_N^d = \omega_{N/d}$, we get $Z(N, f) = d \cdot Z(N/d, f')$, where $f' = f/d$ is the quadratic polynomial obtained from $f$ by dividing every coefficient with $d$. This reduces the modulus to $N/d$, and there can be at most $\log_2 N$ many such steps.

This completes the proof of Lemma 3.1. □

## 4 Modulus is a Power of 2

In this section, we deal with the more difficult case when the modulus, which we denote by $q$ here, is a power of 2: $q = 2^k$ for some $k \geq 1$. We note that the property of an element $c \in \mathbb{Z}_q$ being even or odd is well-defined.

**Lemma 4.1.** *Let $q = 2^k$ for some positive integer $k$ and let $f \in \mathbb{Z}[x_1, \ldots, x_n]$ be a quadratic polynomial over $n$ variables $x_1, \ldots, x_n$. Then $Z(q, f)$ can be evaluated in polynomial time.*

*Proof.* If $k = 1$, $Z(q, f)$ is computable in polynomial time according to [13], so we assume $k > 1$. The algorithm goes as follows. For each round, we can, in polynomial time, either

1. output the correct value of $Z(q, f)$; or

2. construct a new quadratic polynomial $g \in \mathbb{Z}_{q/2}[x_1, \ldots, x_n]$ and reduce the computation of $Z(q, f)$ to the computation of $Z(q/2, g)$; or

3. construct a new quadratic polynomial $g \in \mathbb{Z}_q[x_1, \ldots, x_{n-1}]$, and reduce the computation of $Z(q, f)$ to the computation of $Z(q, g)$.

This gives us a polynomial-time algorithm for evaluating $Z(q, f)$, since we know how to solve the two base cases when either $k = 1$ or $n = 0$ efficiently.

Suppose we have a polynomial $f \in \mathbb{Z}_q[x_1, \ldots, x_n]$ as in (3). Our first step is to transform $f$ so that all the coefficients of its cross terms ($c_{i,j}$, where $1 \leq i < j \leq n$) and linear terms ($c_i$) are even. Assume $f$ does not yet have this property. We let $t$ be the smallest index in $[n]$ such that one of $\{c_t, c_{t,j} : j > t\}$ is odd. By separating out the terms involving $x_t$, we rewrite $f$ as follows

$$f = c_{t,t} \cdot x_t^2 + x_t \cdot f_1(x_1, \ldots, \widehat{x}_t, \ldots, x_n) + f_2(x_1, \ldots, \widehat{x}_t, \ldots, x_n), \tag{5}$$

where $f_1$ is an affine linear function and $f_2$ is a quadratic polynomial. Both $f_1$ and $f_2$ here are over variables $\{x_1, \ldots, x_n\} - \{x_t\}$. (The notation $\widehat{x}_t$ means $x_t$ does not appear in the polynomial.) We let

$$f_1(x_1, \ldots, \widehat{x}_t, \ldots, x_n) = \sum_{i<t} c_{i,t} x_i + \sum_{j>t} c_{t,j} x_j + c_t. \tag{6}$$

By the minimality of $t$, $c_{i,t}$ is even for all $i < t$, and at least one of $\{c_t, c_{t,j} : j > t\}$ is odd.

We claim that

$$Z(q, f) = \sum_{x_1, \ldots, x_n \in \mathbb{Z}_q} \omega_q^{f(x_1, \ldots, x_n)} = \sum_{\substack{x_1, \ldots, x_n \in \mathbb{Z}_q \\ f_1(x_1, \ldots, \widehat{x}_t, \ldots, x_n) \equiv 0 \bmod 2}} \omega_q^{f(x_1, \ldots, x_n)}. \tag{7}$$

This is because

$$\sum_{\substack{x_1, \ldots, x_n \in \mathbb{Z}_q \\ f_1 \equiv 1 \bmod 2}} \omega_q^{f(x_1, \ldots, x_n)} = \sum_{\substack{x_1, \ldots, \widehat{x}_t, \ldots, x_n \in \mathbb{Z}_q \\ f_1 \equiv 1 \bmod 2}} \sum_{x_t \in \mathbb{Z}_q} \omega_q^{c_{t,t} x_t^2 + x_t f_1 + f_2}.$$

However, for any fixed $x_1, \ldots, \widehat{x}_t, \ldots, x_n$, the inner sum over $x_t$ is equal to $\omega_q^{f_2}$ times

$$\sum_{x_t \in [0:2^{k-1}-1]} \omega_q^{c_{t,t} x_t^2 + x_t f_1} + \omega_q^{c_{t,t}(x_t + 2^{k-1})^2 + (x_t + 2^{k-1}) f_1} = \left(1 + (-1)^{f_1}\right) \sum_{x_t \in [0:2^{k-1}-1]} \omega_q^{c_{t,t} x_t^2 + x_t f_1} = 0,$$

since $f_1 \equiv 1 \bmod 2$. Note that we used $(x + 2^{k-1})^2 \equiv x^2 \pmod{2^k}$ when $k > 1$ in the first equation.

Recall that $f_1$ (see (6)) is an affine linear form over $\{x_1, \ldots, \widehat{x}_t, \ldots, x_n\}$. Also note that $c_{i,t}$ is even for all $i < t$, and one of $\{c_t, c_{t,j} : j > t\}$ is odd. We consider the following two cases.

In the first case, $c_{t,j}$ is even for all $j > t$ and $c_t$ is odd, then for any assignment $(x_1, \ldots, \widehat{x}_t, \ldots, x_n)$ in $\mathbb{Z}_q^{n-1}$, $f_1$ is odd. As a result, by (7), $Z(q, f)$ is trivially zero.

In the second case, there exists at least one $j > t$ such that $c_{t,j}$ is odd. Let $\ell > t$ be the smallest of such $j$'s. Then we substitute the variable $x_\ell$ in $f$ with a new variable $x'_\ell$, where (as $c_{t,\ell}$ is odd, $c_{t,\ell}$ is invertible in $\mathbb{Z}_q$)

$$x_\ell = c_{t,\ell}^{-1}\left(2x'_\ell - \left(\sum_{i<t} c_{i,t} x_i + \sum_{j>t, j \neq \ell} c_{t,j} x_j + c_t\right)\right). \tag{8}$$

and let $f'$ denote the new quadratic polynomial in $\mathbb{Z}_q[x_1, \ldots, x_{\ell-1}, x'_\ell, x_{\ell+1}, \ldots, x_n]$. We claim that

$$Z(q, f') = 2 \cdot Z(q, f) = 2 \cdot \sum_{\substack{x_1, \ldots, x_n \in \mathbb{Z}_q \\ f_1 \equiv 0 \bmod 2}} \omega_q^{f(x_1, \ldots, x_n)}.$$

To prove this, we define the following map from $\mathbb{Z}_q^n$ to $\mathbb{Z}_q^n$: $(x_1, \ldots, x'_\ell, \ldots, x_n) \mapsto (x_1, \ldots, x_\ell, \ldots, x_n)$, where $x_\ell$ satisfies (8). It is easy to check that the range of the map is the set of $(x_1, \ldots, x_\ell, \ldots, x_n)$ in $\mathbb{Z}_q^n$ such that $f_1$ is even. Moreover for every such tuple $(x_1, \ldots, x_\ell, \ldots, x_n)$ the number of its preimages in $\mathbb{Z}_q^n$ is exactly 2. The claim then follows.

As a result, to compute $Z(q, f)$, we only need to compute $Z(q, f')$, and the advantage of the new polynomial $f'$ over $f$ is the following property.

**Property 4.1.** *For every cross and linear term that involves $x_1, \ldots, x_t$, its coefficient in $f'$ is even.*

*Proof.* We show that after substituting the variable $x_\ell$ in $f$ with a new variable $x'_\ell$ as in equation (8), the new polynomial $f' \in \mathbb{Z}_q[x_1, \ldots, x_{\ell-1}, x'_\ell, x_{\ell+1}, \ldots, x_n]$ we get satisfies Property 4.1.

To see this, we divide the terms of $f'$ (that we are interested in, i.e. any cross term or linear term that contains at least one of $x_1, \ldots, x_t$) into three groups: cross and linear terms that involve $x_t$; linear terms $x_s$, $s < t$; and cross terms of the form $x_s x_{s'}$, where $s < s'$ and $s < t$.

Firstly, we consider the expression (5) of $f$ after the substitution. The first term $c_{t,t} x_t^2$ remains the same; the second term $x_t f_1$ becomes $2 x_t x'_\ell$ by (8); and $x_t$ does not appear in the third term even after the substitution. Therefore, Property 4.1 holds for $x_t$.

Secondly, we consider the coefficient $c'_s$ of the linear term $x_s$ in $f'$, where $s < t$. Only the following terms in $f$ can possibly contribute to $c'_s$:

$$c_s x_s, \quad c_{\ell,\ell} x_\ell^2, \quad c_{s,\ell} x_s x_\ell, \quad \text{and} \quad c_\ell x_\ell.$$

By the minimality of $t$ both $c_s$ and $c_{s,\ell}$ are even. For $c_{\ell,\ell} x_\ell^2$ and $c_\ell x_\ell$, although we do not know whether $c_{\ell,\ell}$ and $c_\ell$ are even or odd, we know that the coefficient $-c_{t,\ell}^{-1} c_{s,t}$ of $x_s$ in (8) is even since $c_{s,t}$ is even. As a result, for every term in the list above, its contribution to $c'_s$ is even and thus, $c'_s$ is even.

Finally, we consider the coefficient $c'_{s,s'}$ of the term $x_s x_{s'}$ in $f'$, where $s < s'$ and $s < t$. Similarly, only the following terms in $f$ can possibly contribute to $c'_{s,s'}$: (Here we consider the general case when $s' \neq \ell$. The special case when $s' = \ell$ is easier.)

$$c_{s,s'} x_s x_{s'}, \quad c_{\ell,\ell} x_\ell^2, \quad c_{s,\ell} x_s x_\ell, \quad \text{and} \quad c_{\ell,s'} x_\ell x_{s'} \text{ (or } c_{s',\ell} x_{s'} x_\ell).$$

By the minimality of $t$, both $c_{s,s'}$ and $c_{s,\ell}$ are even. Moreover, the coefficient $-c_{t,\ell}^{-1} c_{s,t}$ of $x_s$ in (8) is even. As a result, for every term listed above, its contribution to $c'_{s,s'}$ is even and thus, $c'_{s,s'}$ is even.

This finishes the proof of Property 4.1. □

To summarize, after substituting $x_\ell$ with $x'_\ell$ using (8), we get a new quadratic polynomial $f'$ such that $Z(q, f') = 2 \cdot Z(q, f)$ and for any cross and linear term that involves $x_1, \ldots, x_t$, its coefficient in $f'$ is even. We can repeat this substitution procedure on $f'$: either we show that $Z(q, f')$ is trivially 0, or we get a quadratic polynomial $f''$ such that $Z(q, f'') = 2 \cdot Z(q, f')$ and the parameter $t$ increases by at least one. As a result, given any quadratic polynomial $f$, we can, in polynomial time, either show that $Z(q, f)$ is 0, or construct a new quadratic polynomial $g \in \mathbb{Z}_q[x_1, \ldots, x_n]$ such that $Z(q, f) = 2^d \cdot Z(q, g)$ for some known integer $d \leq n$, and every cross term and linear term of $g$ has an even coefficient. For notational simplicity, we can just assume that the given $f$ in form (3) satisfies this condition. (Or in other words, we rewrite $f$ for $g$.) We will show that, given such a polynomial $f$ in $n$ variables, we can reduce it either to the computation of $Z(q/2, f')$, in which $f'$ is a quadratic polynomial in $n$ variables; or to the computation of $Z(q, f'')$, in which $f''$ is a quadratic polynomial in $n-1$ variables.

We consider the following two cases: $c_{i,i}$ is even for all $i \in [n]$; or at least one of the $c_{i,i}$'s is odd.

In the first case, we know $c_{i,j}$ and $c_i$ are even for all $1 \leq i \leq j \leq n$. We use $c'_{i,j}$ and $c'_i$ to denote integers in $[0 : 2^{k-1} - 1]$ such that $c_{i,j} \equiv 2c'_{i,j} \pmod{q}$ and $c_i \equiv 2c'_i \pmod{q}$, respectively. Then,

$$Z(q, f) = \omega_q^{c_0} \cdot \sum_{x_1, \ldots, x_n \in \mathbb{Z}_q} \omega_q^{2\left(\sum_{i \leq j \in [n]} c'_{i,j} x_i x_j + \sum_{i \in [n]} c'_i x_i\right)} = 2^n \cdot \omega_q^{c_0} \cdot Z(2^{k-1}, f'),$$

where

$$f' = \sum_{i \leq j \in [n]} c'_{i,j} x_i x_j + \sum_{i \in [n]} c'_i x_i$$

is a quadratic polynomial over $\mathbb{Z}_{q/2} = \mathbb{Z}_{2^{k-1}}$. This reduces the computation of $Z(q, f)$ to $Z(q/2, f')$.

In the second case, without loss of generality, we assume $c_{1,1}$ is odd. Then we have

$$f = c_{1,1}(x_1^2 + 2x_1 f_1) + f_2 = c_{1,1}(x_1 + f_1)^2 + f',$$

where $f_1$ is an affine linear form, and $f_2, f'$ are quadratic polynomials, all of which are over $x_2, \ldots, x_n$. We are able to do this because $c_{1,j}$ and $c_1$, for all $j \geq 2$, are even. Now we have

$$Z(q, f) = \sum_{x_1, \ldots, x_n \in \mathbb{Z}_q} \omega_q^{c_{1,1}(x_1 + f_1)^2 + f'} = \sum_{x_2, \ldots, x_n \in \mathbb{Z}_q} \omega_q^{f'} \cdot \sum_{x_1 \in \mathbb{Z}_q} \omega_q^{c_{1,1}(x_1 + f_1)^2} = G(c_{1,1}, q) \cdot Z(q, f').$$

The last equation is because the sum over $x_1 \in \mathbb{Z}_q$ is independent of the value of $f_1$. This reduces the computation of $Z(q, f)$ to $Z(q, f')$ in which $f'$ is a quadratic polynomial in $n-1$ variables.

To sum up, given any quadratic polynomial $f$, we can in polynomial time either output the correct value of $Z(q, f)$; or reduce one of the two parameters, $k$ or $n$, by at least 1. This gives us a polynomial time algorithm to evaluate $Z(q, f)$, when $q$ is a power of 2. □

## 5 #P-Hardness

We first introduce the definition of a *partition function* $Z_\mathbf{A}(\cdot)$ [14, 4, 6, 1, 3], where $\mathbf{A}$ is a symmetric complex matrix. We will give four necessary conditions on the matrix $\mathbf{A}$ for the problem of computing $Z_\mathbf{A}(\cdot)$ being *not* #P-hard. Then we demonstrate the wide applicability of these necessary conditions by reducing $Z_\mathbf{A}(\cdot)$, for some appropriate $\mathbf{A}$, to $Z(N, f)$ and proving that even computing $Z(N, f)$ for some very restricted families of polynomials over a fixed modulus $N$ is #P-hard.

Finally, we show that for a large class of problems defined using $Z(N, f)$, these conditions actually cover all the #P-hard cases. Together with the polynomial-time algorithm presented in Section 3 and 4, they imply an explicit complexity dichotomy theorem for this class.

## 5.1 Partition Functions

Let $\mathbf{A} \in \mathbb{C}^{m \times m}$ be an $m \times m$ symmetric complex matrix. We define the partition function, or graph homomorphism function, $Z_{\mathbf{A}}(\cdot)$ as follows: Given any undirected graph $G = (V, E)$ (here $G$ is allowed to have multi-edges but no self loops)

$$Z_{\mathbf{A}}(G) = \sum_{\xi: V \to [m]} \mathrm{wt}_{\mathbf{A}}(G, \xi), \quad \text{where} \quad \mathrm{wt}_{\mathbf{A}}(G, \xi) = \prod_{(u,v) \in E} A_{\xi(u), \xi(v)}. \tag{9}$$

The computational complexity of $Z_{\mathbf{A}}(\cdot)$, for various $\mathbf{A}$, has been studied intensely [14, 4, 6, 1, 3]. We need the following lemma which can be proved following an important result of Bulatov and Grohe [1]. The proof of the lemma uses the technique of Valiant [16, 15] called interpolation, which is omitted here.

**Lemma 5.1** (The Rank-1 Condition). *Let $\mathbf{A} \in \mathbb{C}^{m \times m}$ be a symmetric matrix, and $\mathbf{A}'$ be the matrix such that $A'_{i,j} = |A_{i,j}|$ for all $i, j$. If there exists a $2 \times 2$ sub-matrix $\mathbf{B}$ of $\mathbf{A}'$ such that $\mathbf{B}$ is of full rank and at least three of the four entries of $\mathbf{B}$ are non-zero, then computing $Z_{\mathbf{A}}(\cdot)$ is #P-hard.*

Next we use Lemma 5.1 to prove a stronger necessary condition for $Z_{\mathbf{A}}(\cdot)$ being *not* #P-hard. The proof can be found in Appendix B. In the statement below, we let $\mathbf{A}_{i,*}$ denote the $i$th row vector of $\mathbf{A}$. We say $\mathbf{A}$ is *M-discrete*, for some integer $M \geq 1$, if every entry of $\mathbf{A}$ is an $M$-th root of unity.

**Lemma 5.2** (Orthogonality). *Let $M$ be a positive integer, and let $\mathbf{A}$ be a symmetric and $M$-discrete $m \times m$ matrix over $\mathbb{C}$. If there exist $i \neq j \in [m]$ such that $\mathbf{A}_{i,*}$ and $\mathbf{A}_{j,*}$ are neither linearly dependent nor orthogonal, then computing $Z_{\mathbf{A}}(\cdot)$ is #P-hard.*

## 5.2 The Group Condition

Next, we prove a much stronger group-theoretic condition for $Z_{\mathbf{A}}(\cdot)$ being not #P-hard, where $\mathbf{A}$ is any *discrete unitary matrix* defined below. A similar condition was first used in the paper by Goldberg et al. [8] for $\{\pm 1\}$ matrices, in the study of $Z_{\mathbf{A}}(\cdot)$ over real matrices $\mathbf{A}$. In the rest of this section, we use $[0 : m-1]$ to index the rows and columns of an $m \times m$ matrix for convenience.

**Definition 5.1** (Discrete Unitary Matrix). *Let $\mathbf{A} \in \mathbb{C}^{m \times m}$ be a symmetric complex matrix. We say $\mathbf{A}$ is an $M$-discrete unitary matrix, for some positive integer $M$, if it is $M$-discrete and satisfies*

— $A_{0,i} = A_{i,0} = 1$ *for all $i \in [0 : m-1]$; and for all $i \neq j \in [0 : m-1]$, $\langle \mathbf{A}_{i,*}, \mathbf{A}_{j,*} \rangle = 0$, where*

$$\langle \mathbf{A}_{i,*}, \mathbf{A}_{j,*} \rangle = \sum_{k=0}^{m-1} \mathbf{A}_{i,k} \overline{\mathbf{A}_{j,k}}.$$

Given two vectors $\mathbf{x}$ and $\mathbf{y} \in \mathbb{C}^m$, we let $\mathbf{x} \circ \mathbf{y}$ denote their Hadamard product: $\mathbf{z} = \mathbf{x} \circ \mathbf{y} \in \mathbb{C}^m$, where $z_i = x_i \cdot y_i$ for all $i$. It is easy to see that operation $\circ$ is associative, and when $\mathbf{A}$ is a discrete unitary matrix, its first row $\mathbf{A}_{0,*} = \mathbf{1}$, the all-1 vector, is an identity element under this operation $\circ$.

**Lemma 5.3** (The Group Condition). *Let $\mathbf{A} \in \mathbb{C}^{m \times m}$ be a symmetric $M$-discrete unitary matrix, for some positive integer $M$. Then $Z_{\mathbf{A}}(\cdot)$ is #P-hard, unless $\mathbf{A}$ satisfies the following* Group Condition:

— *For all $i, j \in [0 : m-1]$, there exists a $k \in [0 : m-1]$ such that $\mathbf{A}_{k,*} = \mathbf{A}_{i,*} \circ \mathbf{A}_{j,*}$.*

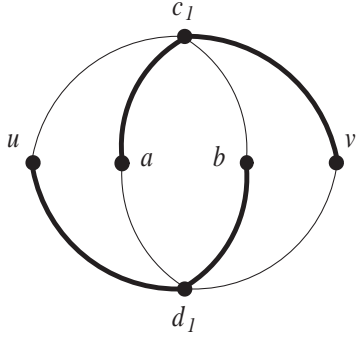

Figure 1: The gadget graph $H^{[p]}$ for $p = 1$, in which the bold line edges represent parallel edges.

**Remark**: This is called the *Group Condition* because if $\mathbf{A}$ satisfies it, then the $m$ row vectors $\mathbf{A}_{i,*}$ of $\mathbf{A}$ form a group under the Hadamard product. To see that, by orthogonality, the map $(i, j) \mapsto k$ where $\mathbf{A}_{k,*} = \mathbf{A}_{i,*} \circ \mathbf{A}_{j,*}$ is uniquely defined. Since entries of $\mathbf{A}$ are nonzero, this product $\circ$ satisfies the cancelation law: $\mathbf{A}_{i,*} \circ \mathbf{A}_{j,*} = \mathbf{A}_{i,*} \circ \mathbf{A}_{j',*} \Longrightarrow \mathbf{A}_{j,*} = \mathbf{A}_{j',*}$, and by orthogonality $j = j'$. Since the row set is finite, by the pigeonhole principle every $\mathbf{A}_{i,*}$ has an inverse. Thus it forms a group under $\circ$.

*Proof.* Assume that $Z_\mathbf{A}(\cdot)$ is not #P-hard. We prove that $\mathbf{A}$ satisfies the Group Condition.

Let $G = (V, E)$ be an undirected graph. For every integer $p \geq 1$, we build a new undirected graph $G^{[p]}$ by replacing every edge $uv \in E$ with a gadget graph called $H^{[p]}$. (cf. the construction used by Goldberg et al. [8].) $H^{[p]}$ has $2p + 4$ vertices $u, v, a, b, c_i, d_i$ where $i \in [p]$. There is one edge between $(u, c_i), (c_i, b), (d_i, a), (d_i, v)$; and $M - 1$ parallel edges between $(c_i, v), (c_i, a), (d_i, b), (d_i, u)$. The gadget for the case $p = 1$ is shown in Figure 1. After replacing every $uv \in E$ with $H^{[p]}$, we get

$$G^{[p]} = (V^{[p]}, E^{[p]}), \quad \text{where } V^{[p]} = V \cup \{a_e, b_e, c_{e,1}, \ldots, c_{e,p}, d_{e,1}, \ldots, d_{e,p} \,|\, e \in E\},$$

and $E^{[p]}$ contains the following edges: For any $e = uv \in E$ and any $i \in [p]$, there is one edge between $(u, c_{e,i}), (c_{e,i}, b_e), (d_{e,i}, a_e), (d_{e,i}, v)$; and $M - 1$ edges between $(c_{e,i}, v), (c_{e,i}, a_e), (d_{e,i}, b_e), (d_{e,i}, u)$.

Now we define, for every $p \geq 1$, the following $m \times m$ matrix $\mathbf{B}^{[p]}$:

$$B^{[p]}_{i,j} = \sum_\xi \mathrm{wt}_\mathbf{A}(H^{[p]}, \xi), \quad \text{for all } i, j \in [0 : m - 1],$$

where the sum is over all possible assignments $\xi$ from the vertex set of $H^{[p]}$ to $[0 : m - 1]$ with $\xi(u) = i$ and $\xi(v) = j$. By the definition of $Z_\mathbf{A}(\cdot)$ and the way we build $G^{[p]}$ from $G$, it can be checked that

$$Z_{\mathbf{B}^{[p]}}(G) = Z_\mathbf{A}(G^{[p]}), \quad \text{for all undirected graphs } G.$$

This gives us a polynomial-time reduction from $Z_{\mathbf{B}^{[p]}}(\cdot)$ to $Z_\mathbf{A}(\cdot)$, so $Z_{\mathbf{B}^{[p]}}(\cdot)$ is also not #P-hard.

Next, we take a closer look at the entries of $\mathbf{B}^{[p]}$. We have

$$B_{i,j}^{[p]} = \sum_{a=0}^{m-1}\sum_{b=0}^{m-1} \left(\sum_{c=0}^{m-1} A_{i,c}\overline{A_{a,c}}A_{b,c}\overline{A_{j,c}}\right)^p \left(\sum_{d=0}^{m-1} \overline{A_{i,d}}A_{a,d}\overline{A_{b,d}}A_{j,d}\right)^p$$

$$= \sum_{a=0}^{m-1}\sum_{b=0}^{m-1}\left|\sum_{c=0}^{m-1} A_{i,c}\overline{A_{a,c}}A_{b,c}\overline{A_{j,c}}\right|^{2p} = \sum_{a=0}^{m-1}\sum_{b=0}^{m-1}\left|\langle \mathbf{A}_{i,*}\circ\overline{\mathbf{A}_{j,*}}, \mathbf{A}_{a,*}\circ\overline{\mathbf{A}_{b,*}}\rangle\right|^{2p}.$$

In the first equation, we used the fact that $(A_{a,c})^{M-1} = \overline{A_{a,c}}$ since $A_{a,c}$ is an $M$-th root of unity. Also note that $\mathbf{B}^{[p]}$ is a symmetric non-negative matrix. Actually, every entry of $\mathbf{B}^{[p]}$ is positive (by taking $a = i$ and $b = j$). As a result, we can apply Lemma 5.1 on $\mathbf{B}^{[p]}$. Since we assumed that $Z_{\mathbf{B}^{[p]}}(\cdot)$ is not #P-hard, we know that $\mathbf{B}^{[p]}$ is of rank 1, for every $p \geq 1$.

For the special case when $j = i \in [0:m-1]$, we have

$$B_{i,i}^{[p]} = \sum_{a=0}^{m-1}\sum_{b=0}^{m-1}\left|\langle \mathbf{1}, \mathbf{A}_{a,*}\circ\overline{\mathbf{A}_{b,*}}\rangle\right|^{2p} = \sum_{a=0}^{m-1}\sum_{b=0}^{m-1}\left|\langle \mathbf{A}_{a,*}, \mathbf{A}_{b,*}\rangle\right|^{2p} = m\cdot m^{2p},$$

since $\mathbf{A}$ is an $M$-discrete unitary matrix. On the other hand, because the rank of $\mathbf{B}^{[p]}$ is 1, we have

$$B_{i,j}^{[p]} = m^{2p+1}, \quad \text{for all } i,j \in [0:m-1] \text{ and all } p \geq 1. \tag{10}$$

Now we use (10) to show that $\mathbf{A}$ satisfies the Group Condition. We start with some notation. Let

$$X_{i,j} = \left\{ \left|\langle \mathbf{A}_{i,*}\circ\overline{\mathbf{A}_{j,*}}, \mathbf{A}_{a,*}\circ\overline{\mathbf{A}_{b,*}}\rangle\right| \;\middle|\; a,b \in [0:m-1]\right\}, \quad \text{for } i,j \in [0:m-1].$$

Clearly $X_{i,j}$ is a finite set for all $i,j$, with cardinality at most $m^2$. Every $x \in X_{i,j}$ satisfies $0 \leq x \leq m$. For each $x \in X_{i,j}$, we let $s_{i,j}(x)$ denote the number of pairs $(a,b) \in [0:m-1] \times [0:m-1]$ such that

$$\left|\langle \mathbf{A}_{i,*}\circ\overline{\mathbf{A}_{j,*}}, \mathbf{A}_{a,*}\circ\overline{\mathbf{A}_{b,*}}\rangle\right| = x.$$

We can now rewrite $B_{i,j}^{[p]}$ as

$$B_{i,j}^{[p]} = \sum_{x \in X_{i,j}} s_{i,j}(x) \cdot x^{2p}, \tag{11}$$

which is equal to $m^{2p+1}$ for all $p \geq 1$. Also note that $s_{i,j}(x)$, for all $x \in X_{i,j}$, do not depend on $p$, and

$$\sum_{x \in X_{i,j}} s_{i,j}(x) = m^2. \tag{12}$$

We can now view equations (11-12) as a linear system in the unknowns $s_{i,j}(x)$. Fix $i,j$, then there are $|X_{i,j}|$ many variables $s_{i,j}(x)$, one for each distinct value $x \in X_{i,j}$. Equations in (11) are indexed by $p \geq 1$. If we choose (12) and (11) for $p = 1, ..., |X_{i,j}| - 1$, this linear system forms an $|X_{i,j}| \times |X_{i,j}|$ Vandermonde matrix which has full rank $|X_{i,j}|$. Also notice that by setting $(a,b) = (i,j)$ and $(a,b) = (i',j)$, where $i' \neq i$, respectively, we get that $m \in X_{i,j}$ and $0 \in X_{i,j}$, respectively. Moreover, we have $s_{i,j}(0) = m^2 - m$, $s_{i,j}(m) = m$, and all other $s_{i,j}(x) = 0$ is a solution to the linear system. Therefore, this must be the unique solution. As a result, we have $X_{i,j} = \{0, m\}$,

$$s_{i,j}(m) = m \quad \text{and} \quad s_{i,j}(0) = m^2 - m, \quad \text{for all } i,j \in [0:m-1].$$

This implies that for all $i, j, a, b \in [0 : m-1]$, $|\langle \mathbf{A}_{i,*} \circ \overline{\mathbf{A}_{j,*}}, \mathbf{A}_{a,*} \circ \overline{\mathbf{A}_{b,*}} \rangle|$ is either $m$ or $0$.

Finally, we prove the Group Condition. Set $j = 0$. Because $\mathbf{A}_{0,*}$ is the all-1 vector, we have

$$\left|\langle \mathbf{A}_{i,*} \circ \mathbf{1}, \mathbf{A}_{a,*} \circ \overline{\mathbf{A}_{b,*}} \rangle\right| = \left|\langle \mathbf{A}_{i,*} \circ \mathbf{A}_{b,*}, \mathbf{A}_{a,*} \rangle\right| \in \{0, m\}, \quad \text{for all } i, a, b \in [0 : m-1].$$

As $\{\mathbf{A}_{a,*} : a \in [0 : m-1]\}$ is an orthogonal basis, where each $\|\mathbf{A}_{a,*}\|^2 = m$, by Parseval, we have

$$\sum_a \left|\langle \mathbf{A}_{i,*} \circ \mathbf{A}_{b,*}, \mathbf{A}_{a,*} \rangle\right|^2 = m \cdot \|\mathbf{A}_{i,*} \circ \mathbf{A}_{b,*}\|^2.$$

Since every entry of $\mathbf{A}_{i,*} \circ \mathbf{A}_{b,*}$ is a root of unity, $\|\mathbf{A}_{i,*} \circ \mathbf{A}_{b,*}\|^2 = m$. Hence

$$\sum_a \left|\langle \mathbf{A}_{i,*} \circ \mathbf{A}_{b,*}, \mathbf{A}_{a,*} \rangle\right|^2 = m^2.$$

As a result, for all $i, b \in [0 : m-1]$, there exists a unique $a$ such that $|\langle \mathbf{A}_{i,*} \circ \mathbf{A}_{b,*}, \mathbf{A}_{a,*} \rangle| = m$.

Since $\mathbf{A}$ is discrete unitary, every entry of $\mathbf{A}_{i,*}$, $\mathbf{A}_{b,*}$ and $\mathbf{A}_{a,*}$ is a root of unity. The inner product $\langle \mathbf{A}_{i,*} \circ \mathbf{A}_{b,*}, \mathbf{A}_{a,*} \rangle$ is a sum of $m$ terms each of complex norm 1. However, to sum to a complex number of norm $m$, every term must be a complex number of unit norm with the *same* argument: they are the same complex number $e^{i\theta}$. Thus, there exists a complex number $e^{i\theta}$, such that $\mathbf{A}_{i,*} \circ \mathbf{A}_{j,*} = e^{i\theta} \cdot \mathbf{A}_{k,*}$. We assert that in fact $e^{i\theta} = 1$, and $\mathbf{A}_{i,*} \circ \mathbf{A}_{b,*} = \mathbf{A}_{a,*}$, since $\mathbf{A}_{i,1} = \mathbf{A}_{b,1} = \mathbf{A}_{a,1} = 1$. This finishes the proof of the lemma. □

## 5.3 Application to the #P-Hardness of $Z(N, f)$

The three necessary conditions we proved in Section 5.2 are very powerful, and can be used to prove the #P-hardness of $Z(N, f)$, for some very restricted families of polynomials over a fixed modulus $N$. We would like to say, e.g., evaluating $Z(N, f)$, when $f$ contains terms $x_1 x_2 x_3$, is #P-hard. However, we have to be very careful; such complexity-theoretic statements are only meaningful for a sequence of polynomials, and not an individual polynomial. This motivates the following definitions.

Let $h \in \mathbb{Z}[x_1, \ldots, x_r]$ be a fixed polynomial (e.g., $h = x_1 x_2 x_3$, where $r = 3$). We say $f \in \mathbb{Z}[x_1, \ldots, x_n]$ is an *h-type* polynomial, if there exists an $r$-uniform hypergraph $G = (V, E)$, where $V = [n]$, such that (Here we allow $G$ to have multiple edges, i.e., $E$ is a multiset; and edges in $E$ are *ordered* subsets of $[n]$ of cardinality $r$)

$$f(x_1, \ldots, x_n) = \sum_{(i_1, \ldots, i_r) \in E} h(x_{i_1}, \ldots, x_{i_r}). \tag{13}$$

**Definition 5.2.** *Let $q = p^t$ be a prime power, and $h \in \mathbb{Z}[x_1, \ldots, x_r]$ be a polynomial. We use $\mathcal{S}[q, h]$ to denote the following problem: Given any $r$-uniform hypergraph $G$, compute $Z(q, f)$, where $f$ is the $h$-type polynomial defined by $G = (V, E)$ using (13).*

First, we use the rank-1 condition to prove the hardness of $\mathcal{S}[q, h_1]$, where $h_1(x_1, x_2, x_3) = x_1 x_2 x_3$.

**Corollary 5.1.** *For any fixed prime power $q = p^t$, $\mathcal{S}[q, h_1]$ is #P-hard.*

*Proof.* Let $\mathbf{A}$ be the following $q \times q$ complex matrix (here we use $[0 : q-1]$ to index $\mathbf{A}$):

$$A_{i,j} = \sum_{k \in [0:q-1]} \omega_q^{ijk}, \quad \text{for all } i, j \in [0 : q-1].$$

It is clear that $\mathbf{A}$ is symmetric. Moreover, we have $A_{0,0} = A_{0,1} = A_{1,0} = q$ but $A_{1,1} = 0$. As a result, by Lemma 5.1 the problem of computing $Z_{\mathbf{A}}(\cdot)$ is #P-hard. However, there is also a polynomial-time reduction from $Z_{\mathbf{A}}(\cdot)$ to $\mathcal{S}[q, h_1]$. Given any input graph $G = (V, E)$ of $Z_{\mathbf{A}}(\cdot)$, we build a 3-uniform hypergraph $G' = (V', E')$ as follows:

$$V' = V \cup \{w_e : e \in E\} \quad \text{and} \quad E' = \{(u, v, w_e) : e = uv \in E\}.$$

By definition, it is easy to check that $Z_{\mathbf{A}}(G) = Z(q, f)$, where $f$ is the $h_1$-type polynomial defined by $G'$. As a result, $\mathcal{S}[q, h_1]$ is also #P-hard. □

Second, we show how to employ the Orthogonality condition to prove the #P-hardness of $\mathcal{S}[q, h_2]$ and $\mathcal{S}[q, h_3]$, where $h_2(x_1, x_2) = x_1^2 x_2$ and $h_3(x_1, x_2) = x_1 x_2 + x_1^2 x_2^2$, respectively.

**Corollary 5.2.** *For any prime power $q \notin \{2, 4\}$, $\mathcal{S}[q, h_2]$ is #P-hard; and for any odd prime power $q$, $\mathcal{S}[q, h_3]$ is #P-hard.*

*Proof.* Let $\mathbf{A}$ be the following $q \times q$ matrix defined by $h_2$:

$$A_{i,j} = \omega_q^{h_2(i,j) + h_2(j,i)}, \quad \text{for all } i, j \in [0 : q-1],$$

and we show that $Z_{\mathbf{A}}(\cdot)$ can be reduced to $\mathcal{S}[q, h_2]$. Given any undirected graph $G = (V, E)$, we build the following directed graph $G' = (V, E')$, where $E' = \{(u, v), (v, u) : uv \in E\}$. Then we always have $Z_{\mathbf{A}}(G) = Z(q, f)$, where $f$ is the $h_2$-type polynomial defined by $G'$. It is easy to check that if $q$ is odd, then $\mathbf{A}_{0,*}$ and $\mathbf{A}_{1,*}$ are neither linearly dependent nor orthogonal; and if $q \notin \{2, 4\}$ is a power of 2, then $\mathbf{A}_{0,*}$ and $\mathbf{A}_{2,*}$ are neither linearly dependent nor orthogonal. This proves the #P-hardness of $\mathcal{S}[q, h_2]$.

For $\mathcal{S}[q, h_3]$, one can similarly define the following matrix $\mathbf{A}$:

$$A_{i,j} = \omega_q^{h_3(i,j)}, \quad \text{for all } i, j \in [0 : q-1], \tag{14}$$

and prove that $Z_{\mathbf{A}}(\cdot)$ can be reduced to $\mathcal{S}[q, h_3]$. Moreover, when $q$ is an odd prime power, the two vectors $\mathbf{A}_{0,*}$ and $\mathbf{A}_{1,*}$ are neither linearly dependent nor orthogonal. This finishes the proof. □

However, even with the Orthogonality condition, we are not able to show the hardness of $\mathcal{S}[q, h_3]$ when $q = 2^t$ and $t \geq 3$. This is because the matrix $\mathbf{A}$ defined by $h_3$ in (14) is discrete unitary and thus, passes the orthogonality test. To prove its #P-hardness, we need to use the much stronger Group condition:

**Corollary 5.3.** *For any $q = 2^t \notin \{2, 4\}$, $\mathcal{S}[q, h_3]$ is #P-hard.*

*Proof.* When $q = 2^t \notin \{2, 4\}$, the matrix $\mathbf{A}$ defined by $h_3$ is $q$-discrete unitary but does not satisfy the Group Condition. The #P-hardness of $\mathcal{S}[q, h_3]$ then follows from Lemma 5.3. □

## 5.4 A Dichotomy Theorem for $\mathcal{S}[q, h]$

Let $q$ be a prime power, and $h \in \mathbb{Z}_q[x_1, x_2]$ be a symmetric polynomial. By the proofs of Corollary 5.2 and 5.3 the problem $\mathcal{S}[q, h]$ is computationally equivalent to $Z_{\mathbf{A}}(\cdot)$ where $\mathbf{A}$ is the following $q \times q$ and $q$-discrete matrix:

$$A_{i,j} = \omega_q^{h(i,j)}, \quad \text{for all } i,j \in [0:q-1]. \tag{15}$$

Although the *Orthogonality* condition and the *Group* condition can be used to prove the #P-hardness of $\mathcal{S}[q,h]$ for many interesting polynomials $h$, as demonstrated in Corollary 5.2 and 5.3, it does not cover all the #P-hard problems $\mathcal{S}[q,h]$. For example, even if we assume that $h$ is symmetric and every monomial in $h(x_1,x_2)$ contains both $x_1$ and $x_2$ (and thus, $h(0,x) = h(x,0) = 0$ for all $x \in \mathbb{Z}_q$ and the matrix $\mathbf{A}$ defined in (15) is symmetric and *normalized*: $A_{0,i} = A_{i,0} = 1$ for all $i$), the Group condition cannot deal with the case when there exist indices $i \neq j \in [0:q-1]$ such that $\mathbf{A}_{i,*} = \mathbf{A}_{j,*}$. We will use $\mathcal{C}$ to denote this class of problems.

We can prove a stronger theorem — the fourth condition, which is a *strengthening* of the current Group condition, leading to a complexity dichotomy theorem for $\mathcal{C}$. The proof is omitted here.

**Lemma 5.4** (The Generalized Group Condition). *Let $\mathbf{A}$ be any $m \times m$ symmetric, normalized, and $M$-discrete matrix for some positive integer $M$, such that for all indices $i$ and $j \in [0:m-1]$, either $\mathbf{A}_{i,*} = \mathbf{A}_{j,*}$ or $\langle \mathbf{A}_{i,*}, \mathbf{A}_{j,*} \rangle = 0$. Let $T_1, \ldots, T_\ell$ be a partition of $[0:m-1]$ such that*

$$\mathbf{A}_{i,*} = \mathbf{A}_{j,*} \iff \exists k \in [\ell] : i,j \in T_k.$$

*Then $Z_\mathbf{A}(\cdot)$ is #P-hard unless $\mathbf{A}$ satisfies the following* Generalized Group Condition:

— $\forall k \in [\ell]$, $|T_k| = m/\ell$; and $\forall i,j \in [0:m-1]$, $\exists k \in [0:m-1]$ such that $\mathbf{A}_{k,*} = \mathbf{A}_{i,*} \circ \mathbf{A}_{j,*}$.

By combining the Generalized Group Condition with the Orthogonality Condition, we are able to show that for every problem $\mathcal{S}[q,h] \in \mathcal{C}$, either $\mathcal{S}[q,h]$ is #P-hard; or we have $\mathbf{A} = \mathbf{J} \otimes \mathbf{A}'$ where $\mathbf{J}$ is an all-1 matrix and $\mathbf{A}'$ is a $q$-discrete unitary matrix that satisfies the original Group Condition. The latter can ultimately lead to a polynomial-time algorithm for $Z_\mathbf{A}(\cdot)$ as well as $\mathcal{S}[q,h]$, using the algorithm developed in Section 3 and 4.

## Acknowledgments

We would like to thank Eric Bach, Richard Brualdi, Michael Kowalczyk, Endre Szemeredi and Mingji Xia for helpful discussions.

# References

[1] A. Bulatov and M. Grohe. The complexity of partition functions. *Theoretical Computer Science*, 348(2):148–186, 2005.

[2] J.-Y. Cai, X. Chen, and P. Lu. Graph homomorphisms with complex values: A dichotomy theorem. In *Proceedings of the 37th International Colloquium on Automata, Languages and Programming*, 2010.

[3] M.E. Dyer, L.A. Goldberg, and M. Paterson. On counting homomorphisms to directed acyclic graphs. *Journal of the ACM*, 54(6), 2007.

[4] M.E. Dyer and C. Greenhill. The complexity of counting graph homomorphisms. In *Proceedings of the 9th International Conference on Random Structures and Algorithms*, pages 260–289, 2000.

[5] A. Ehrenfeucht and M. Karpinski. The computational complexity of (XOR, AND)-counting problems. *University of Bonn*, (Technical Report 8543-CS), 1990.

[6] M. Freedman, L. Lovász, and A. Schrijver. Reflection positivity, rank connectivity, and homomorphism of graphs. *Journal of the American Mathematical Society*, 20:37–51, 2007.

[7] J. von zur Gathen and J. Gerhard. *Modern Computer Algebra*. Cambridge University Press, 2003.

[8] L.A. Goldberg, M. Grohe, M. Jerrum, and M. Thurley. A complexity dichotomy for partition functions with mixed signs. In *Proceedings of the 26th International Symposium on Theoretical Aspects of Computer Science*, pages 493–504, 2009.

[9] J. Grabmeier, E. Kaltofen, and V. Weispfenning. *Computer Algebra Handbook*. Springer, 2003.

[10] L. Hua. *Introduction to Number Theory*. Springer-Verlag, 1982.

[11] K. Ireland and M. Rosen. *A Classical Introduction to Modern Number Theory*. Springer, 1998.

[12] S. Lang. *Algebraic Number Theory*. Addison-Wesley, 1970.

[13] R. Lidl and H. Niederreiter. *Finite Fields*. volume 20 of Encyclopedia of Mathematics and its Applications. Cambridge University Press, Cambridge, 1997.

[14] L. Lovász. Operations with structures. *Acta Mathematica Hungarica*, 18:321–328, 1967.

[15] L.G. Valiant. The complexity of computing the permanent. *Theoretical Computer Science*, 8:189–201, 1979.

[16] L.G. Valiant. The complexity of enumeration and reliability problems. *SIAM Journal on Computing*, 8(3):410–421, 1979.

# Appendix

## A  Gauss Sums

In this section, we gather together some facts we need about Gauss sums. More information can be found in the book "Algebraic Number Theory" by Serge Lang [12].

Let $a$ and $b$ be non-zero integers, $b > 0$, and $\gcd(a, b) = 1$. Let $G(a, b)$ denote the following sum:

$$G(a, b) = \sum_{x \in \mathbb{Z}_b} \omega_b^{ax^2}.$$

The following identities are known.

- $G(a, 1) = 1$.

- If $p$ is an odd prime, then

$$G(a, p) = \left(\frac{a}{p}\right) \cdot G(1, p), \quad \text{where } \left(\frac{a}{p}\right) \text{ is the Legendre symbol.}$$

- If $p$ is an odd prime, and $r \geq 2$ is an integer, then $G(a, p^r) = p \cdot G(a, p^{r-2})$.

- Let $b, c \geq 1$, $\gcd(b, c) = 1$, and $\gcd(a, bc) = 1$. Then $G(a, bc) = G(ab, c) \cdot G(ac, b)$.

- If $b$ is odd $\geq 1$, then

$$G(a, b) = \left(\frac{a}{b}\right) \cdot G(1, b), \quad \text{where } \left(\frac{a}{b}\right) \text{ is the Jacobi symbol.}$$

- When $a = 1$, we have (Note the positive sign in front of $\sqrt{b}$.)

$$G(1, b) = \begin{cases} (1+i)\sqrt{b} & \text{if } b \equiv 0 \pmod{4} \\ \sqrt{b} & \text{if } b \equiv 1 \pmod{4} \\ 0 & \text{if } b \equiv 2 \pmod{4} \\ i\sqrt{b} & \text{if } b \equiv 3 \pmod{4}. \end{cases}$$

- For odd $a$,

$$G(a, 2^r) = \left(\frac{-2^r}{a}\right) \cdot \epsilon(a) \cdot G(1, 2^r), \quad \text{where } \left(\frac{-2^r}{a}\right) \text{ is the Jacobi symbol,}$$

and

$$\epsilon(a) = \begin{cases} 1 & \text{if } a \equiv 1 \pmod{4} \\ i & \text{if } a \equiv 3 \pmod{4}. \end{cases}$$

An immediate conclusion is that the Gauss sum $G(a, b)$ can be computed in polynomial time, in input length $\log a + \log b$, without knowing their prime factorizations.

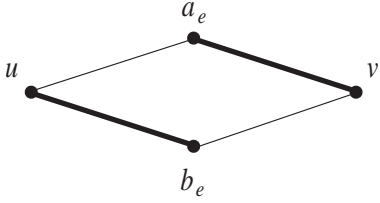

Figure 2: The gadget graph $H$, in which the bold line edges represent parallel edges.

## B  Proof of Lemma 5.2

Let $G = (V, E)$ be an undirected graph. We construct a new undirected graph $G^*$ by replacing each edge $uv \in E$ with a gadget graph $H$ as shown in Figure 2. $H$ has four vertices $u, v, a, b$. There is one edge between $(u, a)$ and $(b, v)$, and $M - 1$ parallel edges between $(a, v)$ and $(u, b)$.

More exactly, we define the new graph $G^* = (V^*, E^*)$ as follows:

$$V^* = V \cup \{a_e, b_e \,|\, e \in E\}$$

and $E^*$ contains exactly the following edges: For each $e = uv \in E$,

1. one edge between $(u, a_e)$ and $(b_e, v)$; and
2. $(q - 1)$ parallel edges between $(a_e, v)$ and $(u, b_e)$.

Now we define $\mathbf{A}^*$ as the following $M \times M$ symmetric matrix:

$$A^*_{i,j} = \sum_{a,b \in [M]} A_{i,a} A_{j,b} \overline{A_{j,a} A_{i,b}} = \left| \sum_{a \in [M]} A_{i,a} \overline{A_{j,a}} \right|^2, \quad \text{for all } i, j \in [M].$$

By the definition of $Z_{\mathbf{A}}(\cdot)$ and the way we build $G^*$ from $G$ using $H$, it can be checked that

$$Z_{\mathbf{A}^*}(G) = Z_{\mathbf{A}}(G^*), \quad \text{for all undirected graphs } G.$$

This gives us a polynomial-time reduction from $Z_{\mathbf{A}^*}(\cdot)$ to $Z_{\mathbf{A}}(\cdot)$.

Next we observe the entries of the new matrix $\mathbf{A}^*$. Without loss of generality, we assume the row 1 and row 2 of $\mathbf{A}$ are neither linearly dependent nor orthogonal. As a result, we have

$$A^*_{0,0} = A^*_{1,1} = M^2, \quad \text{but} \quad 0 < A^*_{0,1} = A^*_{1,0} < M^2.$$

As a result, by Lemma 5.1, $Z_{\mathbf{A}^*}(\cdot)$ is #P-hard and thus, $Z_{\mathbf{A}}(\cdot)$ is also #P-hard.